\documentclass[%
onecolumn,pra,
superscriptaddress,
nofootinbib,
 amsmath,amssymb,
 aps,
11pt,
]{revtex4-1}

\usepackage{graphicx}
\usepackage{array}
\usepackage{dcolumn}
\usepackage{bm}
\usepackage{listings}
\usepackage{scalefnt}
\usepackage{xcolor}
\usepackage{multirow}
\usepackage{siunitx}
\usepackage{booktabs}
\usepackage{amsthm,amsfonts}
\usepackage{adjustbox}
\usepackage{tabularx}
\usepackage{capt-of}
\usepackage{lmodern}

\usepackage{algorithm}
\usepackage[noend]{algpseudocode}

\usepackage{enumitem}
\usepackage{tikz}

\usepackage{braket}

\begin{document}

\title{Two-step approach to scheduling quantum circuits}

\author{Gian Giacomo Guerreschi}
\email{Email: gian.giacomo.guerreschi@intel.com}
\affiliation{Parallel Computing Lab, Intel Corporation}
\author{Jongsoo Park}
\affiliation{Parallel Computing Lab, Intel Corporation}
\date{\today}

\begin{abstract}
As the effort to scale up existing quantum hardware proceeds, it becomes necessary to schedule quantum gates in a way that minimizes the number of operations. There are three constraints that have to be satisfied: the order or dependency of the quantum gates in the specific algorithm, the fact that any qubit may be involved in at most one gate at a time, and the restriction that two-qubit gates are implementable only between connected qubits. The last aspect implies that the compilation depends not only on the algorithm, but also on hardware properties like connectivity. Here we suggest a two-step approach in which logical gates are initially scheduled neglecting connectivity considerations, while routing operations are added at a later step in a way that minimizes their overhead. We rephrase the subtasks of gate scheduling in terms of graph problems like edge-coloring and maximum subgraph isomorphism.
While this approach is general, we specialize to a one dimensional array of qubits to propose a routing scheme that is minimal in the number of exchange operations.
As a practical application, we schedule the Quantum Approximate Optimization Algorithm in a linear geometry and quantify the reduction in the number of gates and circuit depth that results from increasing the efficacy of the scheduling strategies.
\end{abstract}

\maketitle


\section{Introduction}
\label{sec:introduction}

Modern computers rely on optimized instruction scheduling to takes full advantage of the computing capability of microchips. Harnessing more parallelism, which can be achieved by scheduling operations on multiple compute units at the same time, greatly contributes to realizing high performance in compute-intensive applications in various domains such as scientific computing, big data analytics, and machine learning.
Compared to the mature field of classical computing, quantum computing is an area of research that only recently has moved into technological relevance \cite{Johnson2011,Devitt2016,Barends2016,Versluis2016a,IBM2017,Sete2016}.
While providing the list of instructions for machines with only a handful of qubits is a relatively simple task, for larger machines the scheduling problem needs to be addressed with systematic methods. Pioneering works have already addressed the synthesis and compilation of quantum circuits either as part of extensive software frameworks \cite{Wecker2014,Smith2016a,Steiger2016} or as an independent problem \cite{Beals2013,Brierley2015a,Martinez2016,Venturelli2018}, but the field remains vastly unexplored.

Consider the three main constraints that a scheduler for quantum algorithms has to take into account: the first one is due to logical dependencies, \emph{i.e.} the order of operations inherent in the algorithm. The other two are due to hardware constraints: no qubit can be involved in more than one gate at the same time and two-qubit gates can be implemented only between qubits that are physically connected or interacting. The last constraint, in particular, implies that routing operations are needed whenever an algorithm requires logical gates between unconnected qubits. The resulting overhead may be large, especially for less connected hardware, and may even affect the overall algorithmic scaling.
We propose a two-step approach in which the logical gates are initially ordered to form the {\em Logical Data Precedence Graph} (LDPG) in a way that neglects the connectivity constraint and aims at identifying the critical path and reducing the overall running time. By extension, this step encourages parallelism on the execution of the logical quantum gates. The necessary routing operations are added in the second phase, and their number is minimized according to heuristic strategies that favor the gates that do not require routing operations.

Our work originates from the realization that compiling quantum circuits is a task that requires specialized algorithms and cannot be efficiently performed when expensive intermediate representations are used. In the following, we describe how the subtasks of gate scheduling can be naturally recast in terms of graph problems like edge-coloring and maximum subgraph isomorphism. We believe that this connection will be particularly fruitful since it allows the solution of well studied problems to have positive impact in quantum computing technologies.

Finally, a concrete implementation of our approach is illustrated by scheduling the Quantum Approximate Optimization Algorithm for a hardware with linear connectivity, arguably the most limiting geometry that still allows for scalable hardware.

\section{Requirements for practical quantum circuits and two-step approach}
\label{sec:two-step}

Optimizing schedules, in both the classical and quantum case, is a hard problem that cannot be exactly solved with polynomial efforts \cite{Ozer1998,graham1966bounds}.
Still, schedulers based on heuristic methods have effectively harnessed the parallelism in classical computers. It is reasonable to expect that a similar benefit is achievable by scheduling quantum algorithms.

In this section, we describe three kinds of constraints that must be satisfied in practical implementations of quantum algorithms. We propose a two-step approach to deal with these constraints as separate sub-problems and show how they can be tackled by specialized schedulers.

According to the gate model of quantum computation, information is stored in the quantum state of the qubit register. To manipulate such information in any conceivable way, it is sufficient to act with operations (called gates) that involve one or, at most, two qubits at a time. At the end of the computation, the information is retrieved by measuring the qubit register%
.

Implementing algorithms on specific hardware may differ based on what set of gates is available, on the number of gates required to decompose multi-qubit operations into gates acting on at most two qubits, and on how efficiently those gates can be error corrected. These aspects are of fundamental importance in the pre-compilation phase of quantum algorithms, but the focus of our work is on the compilation tasks that still lie ahead.

In fact, the pre-compilation phase provides the sequence of single- and two-qubit logical gates that have to be performed in order to execute the algorithm. Once such a sequence is available, those operations have to be scheduled as a list of physical instructions. At this level, there are three types of constraints that must be satisfied:
\begin{description}
	\item[Logical dependency] Certain operations have to be performed after the completion of previous operations. The dependencies are explicitly provided by the quantum algorithm.
	\item[Exclusive activation] A qubit can only be involved in a single operation at a time. This also applies to quantum operations that formally commute: While their temporal order may not matter, they still cannot be executed at the same time if they share a qubit.
	\item[Physical connectivity] Two-qubit gates are possible only between qubits that are connected according to the hardware topology. In physical terms, a connection means that it is possible to generate and control a sufficiently strong interaction between the two qubits.
\end{description}

Notice that the exclusive activation constraint can be expressed differently for different hardware realizations. While we focus on the case described above, an extended formulation may require, for example, that no connected qubits are involved in different gates at the same time (as it seems to be the case for the hardware under development at Google \cite{Boixo2016}).

\begin{figure}[b!]
\centering
\includegraphics[width=0.75\linewidth]{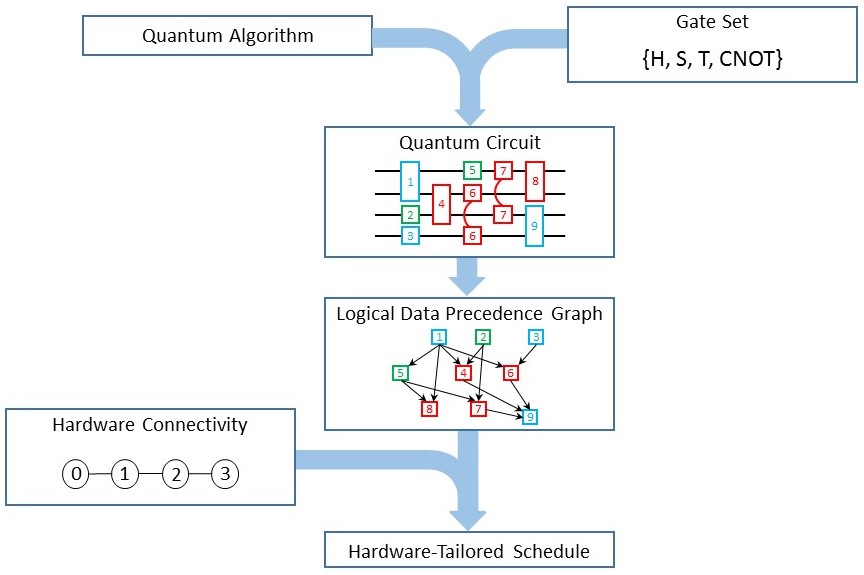}
\caption{Summary of the two-step approach to schedule quantum circuits. Initially, the quantum algorithm must be phrased in terms of a quantum circuit that only involves gates from the (universal) set available to the specific hardware. This is indicated by the two boxes ``quantum algorithm'' and ``gate set'' merging into the ``quantum circuit'': this constitutes the input of our scheduler. The first step is then to capture the logical dependency and order of quantum operations by constructing the Logical Data Precedence Graph. The second step requires the knowledge of the connectivity graph of the particular machine and introduces the overhead due to routing and tie-breaking strategies. The final result is the schedule, provided in terms of the Physical Data Precedence Table.}\label{fig:scheme}
\end{figure}

We propose a two-step approach. In the first step the temporal order determines a ``Logical Data Precedence Graph'' (LDPG) that is used to assign a priority value to each and every gate. In the second step, the exclusive activation and connectivity constraints determine the necessary routing operations that are added to complete the schedule. FIG.~\ref{fig:scheme} provides a pictorial illustration of the process. Finally, the schedule is described by means of the ``Physical Data Precedence Table'' (PDPT).

Using concepts and notation from the standard literature on schedulers \cite{Cooper1998,Ozer1998}, the LDPG can be visualized as a graph $\mathcal{G} = (\mathcal{N},\mathcal{E})$ in which a node $n\in\mathcal{N}$ represents a gate and a directed edge $e=(n_i,n_j)\in\mathcal{E}$ represents a dependency between the quantum operations (in this case, $n_j$ depends on $n_i$). Only explicit dependencies are shown with edges; meaning that, if gate $n_2$ has to follow $n_1$ and gate $n_3$ has to follow $n_2$, we do not explicitly indicate the transitive dependency that $n_3$ has to follow $n_1$.

Additional rules are required to properly take advantage of the fact that certain quantum gates commute and, therefore, their order can be relaxed. In the next section, 
we discuss the rules to create the LDPG and visualize the result in FIG.~\ref{fig:example_LDPG}. From the LDPG, one can assign a priority value to each gate: The basic idea is that the priority of a certain gate increases if more gates logically depend on it. Large priorities characterize the gates that are along the critical path (\emph{i.e.} the most time consuming sequence of logically dependent gates) and, therefore, indicate the operations that should be executed as early as possible.

The PDPT is a table with as many rows as qubits in the hardware and one column for each clock-cycle required to schedule either logical gates or routing operations. The total number of columns then corresponds to the circuit duration in clock-cycles, and the attribute ``physical'' indicates that connectivity constraints are included. We fill the PDPT starting from the gates with the highest priority. If multiple gates have an equal priority, we prioritize gates that satisfy the connectivity constraint from the current mapping between physical and logical qubits. If this is not possible, we propose to derive the order from the solution of the maximum subgraph isomorphism between the connectivity graph and the interaction graph (the latter describing the gates that needs to be scheduled with certain priority). Operations that exchange two connected qubits (also called SWAP gates in the following) are added at this time. Details are provided in section~\ref{sec:PDPT}.

\section{Construction of the Logical Data Precedence Graph with priority values}
\label{sec:LDPG}

The input is a quantum circuit composed by one- and two-qubit gates belonging to the set of gates available in the specific hardware. To preserve generality, the set of gates must be capable of universal quantum computation%
\footnote{To achieve universal quantum computation it is not necessary to implement many kinds of gates. A well-studied case is the discrete set composed by four gates $\{H,T,S,CNOT\}$: Three one-qubit gates and one two-qubit gate suffice to approximate any quantum operation within arbitrary precision.}.
A preliminary step may be needed to associate the quantum algorithm with a circuit of such form: For example, multi-qubit gates or arbitrary single-qubit rotations must be decomposed in terms of the available gates.
The output is the LDPG together with a priority value associated to each node (here representing a gate).

\begin{figure}[b!]
\centering
\includegraphics[width=0.7\linewidth]{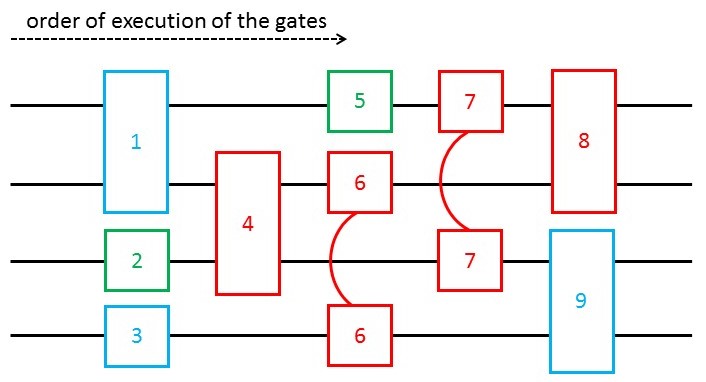}
\caption{Example of a quantum circuit with three groups of pair-wise commuting gates (identified by the same color). Each line corresponds to a qubit, each box to an operation, and the order of execution is from left to right.
While it is clear that gates $n_1$ and $n_7$ are consecutive to gate $n_5$, it is less obvious that also gate $n_8$ is consecutive to $n_5$. This is due to the fact that $n_7$ commutes with $n_8$: had we chosen another order in the picture, $n_5$ and $n_8$ would be in direct sequence also visually. As a second example, gate $n_6$ is consecutive to gates $n_1$, $n_4$, $n_8$, $n_3$, and $n_9$.}\label{fig:consecutive_gates}
\end{figure}

In the following we refer to two gates as ``consecutive'' if they are not separated by another operation and act on, at least, one common qubit. Notice, however, that quantum gates may commute, meaning that their order of execution has no relevance on the combined operation. For this reason, the definition of consecutive gates must be generalized to include gates that appear separated by commuting operations, but that can effectively be brought in direct sequence through gate reordering/commutation only. FIG.~\ref{fig:consecutive_gates} clarifies this definition with an example.

We call ``parents'' (respectively ``children'') of a certain gate the consecutive and non-commuting gates that logically precede (respectively follow) the operation. In FIG.~\ref{fig:consecutive_gates}, gate $n_4$ has two parents $n_1$, $n_2$ and one child $n_9$. While $n_6$ and $n_7$ are consecutive gates for $n_4$, they commute with it and so they do not qualify for parent/child relationships.

Without consecutive gates that commute, each two-qubit operation has at most 2 parents and at most 2 children, while single-qubit operations have unique parent and child. This property is not preserved when commuting gates are present. Referring to FIG.~\ref{fig:consecutive_gates}, gate $n_1$ has 4 children, namely $n_4$, $n_5$, $n_6$, and $n_8$.

\begin{figure}[b!]
\centering
\includegraphics[width=0.55\linewidth]{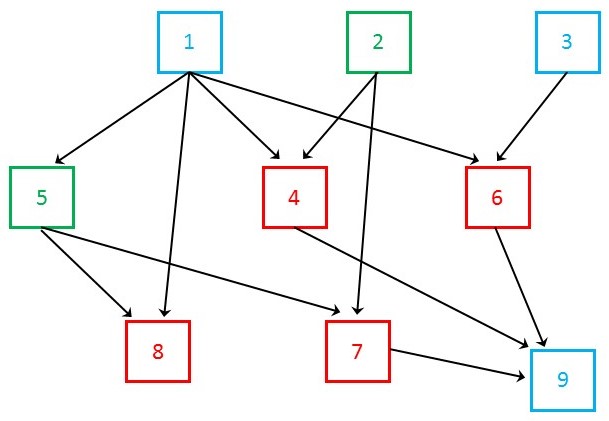}
\caption{Logical Data Precedence Graph corresponding to the quantum circuit in FIG.~\ref{fig:consecutive_gates}. Recall that gates with the same color are pairwise commuting. According to the rules explained in the main text, observe that gate $n_1$ is a parent of $n_8$ since $n_8$ could be reordered ahead of $n_4$ and $n_6$. In contrast, gate $n_1$ is not a parent of $n_7$ since gate $n_5$ separates them.}\label{fig:example_LDPG}
\end{figure}

The parent/child relationship can be used to construct the LDPG efficiently. Each gate depends on its parents and, conversely, its children depend on it. A directed edge is then drawn from each parent to the gate and from the gate to each of its children. For simplicity, quantum operations can be added as nodes of the LDPG starting from the leftmost gates (those that do not have parents), until the rightmost gates (those that do not have children) complete it. FIG.~\ref{fig:example_LDPG} shows the LDPG corresponding to the quantum circuit if FIG.~\ref{fig:consecutive_gates}.

The graph has directed edges, no self-loop and is acyclic. The nodes with only incoming edges and no outgoing edges (i.e. gates without children gates) are called leaves in graph theory, but sometimes we say that they belong to the ``last generation'' following the parent/child relationship.

After an LDPG is built, priorities are assigned to each node in the graph. An effective strategy (common in scheduling algorithms for classical computers)
is to define the priority as the latency-weighted depth of the node. The depth of a node $n_i$ corresponds to the maximum number of nodes traversed along any directed path from $n_i$ to any gate in the last generation (\emph{i.e.} a leaf). Since not all gates require the same time or effort in actual implementation, we associate a latency to each kind of gates. For example, the latency may represent the time required for the physical interaction to generate the specific quantum operation, but it may also depend on the fidelity of the quantum operation itself. Hereafter, we will think of the gate latency as the time required by the corresponding quantum operation and denote it as $t_i=\text{latency}(n_i)$. Latencies must be positive so every gate has a larger priority than its children and lower priority than its parents. In practice, the priority is a scalar value consistent with all logical dependencies.

Defining $\mathcal{P}(n_i)$ as the set of paths connecting node $n_i$ with any of the LDPG leaves, the latency-weighted depth $p_i=\text{priority}(n_i)$ is then computed according to:
\begin{equation}
	p_i = \max_{P\in\mathcal{P}(n_i)} \sum_{n_j\in P} t_j \; ,
\end{equation}
with $P$ being an arbitrary path from $n_i$ to any leaf, and $t_j$ being the latency of node $n_j$ in path $P$.
Priorities can be efficiently computed by traversing the (directed and acyclic) graph in a post-order, starting from the gates in the last generation ($p_j=t_j$ when $n_j\in\,$leaves) and, for each node, adding its latency to the maximum of its children's priorities. A simple, but realistic, situation is obtained when all priorities are unitary. This case represents quantum circuits executed in a synchronous way in which each and every gate takes a fixed amount of time. For the numerical study in section~\ref{sec:QAOA}, we consider such synchronous model together with the simple asynchronous case where $t_i=1$ for one-qubit gates and $t_i=2$ for two-qubit gates. See TABLE~\ref{table:example_priority} and FIG.~\ref{fig:example_priority} for an illustration.

\begin{table}[ht]
\begin{minipage}[b]{0.45\linewidth}
\centering
\begin{tabular}{ | c || c | c |}
\hline
    \hspace{4mm}gate\hspace{4mm} &      priority      &     priority     \\
                                 & (latency always 1) & (latency 1 or 2) \\
\hline \hline
    9 & 1 & 2 \\ \hline
    8 & 1 & 2 \\ \hline
    7 & 2 & 4 \\ \hline
    6 & 2 & 4 \\ \hline
    5 & 3 & 5 \\ \hline
    4 & 2 & 4 \\ \hline
    3 & 3 & 5 \\ \hline
    2 & 3 & 5 \\ \hline
    1 & 4 & 7 \\
\hline
\end{tabular}
\vspace{0.3cm}
\caption{Priority values for the circuit in FIG.~\ref{fig:example_LDPG}. Two cases are considered in which the latency is either always equal to unity or is increased to 2 for two-qubit gates.}
\label{table:example_priority}
\end{minipage}\hfill
\begin{minipage}[b]{0.5\linewidth}
\centering
\includegraphics[width=50mm]{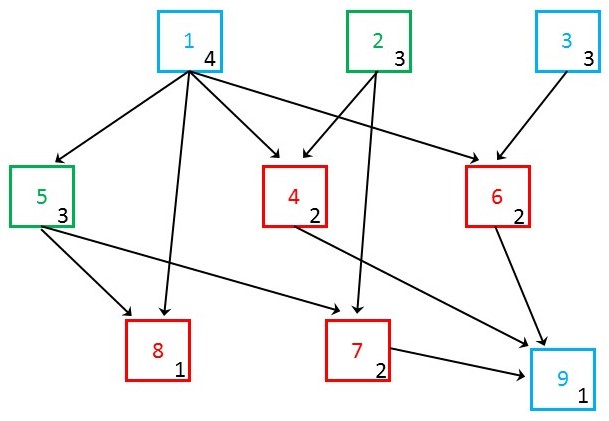}\\
\vspace{0.4cm}
\includegraphics[width=50mm]{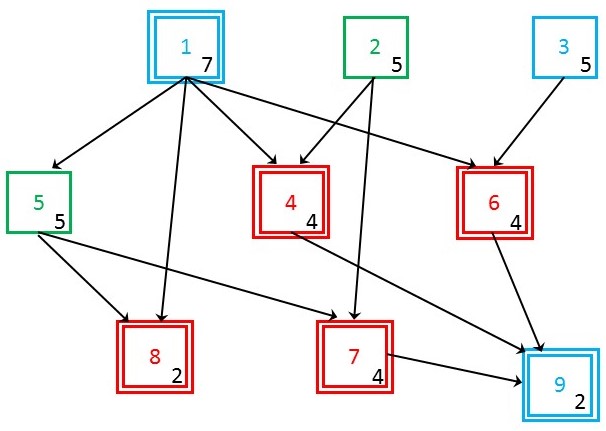}
\captionof{figure}{LDPG including the gate priorities as numbers on the lower right side of the nodes (here depicted like boxes). Gate latencies are indicated by the number of lines forming the corresponding box.}
\label{fig:example_priority}
\end{minipage}
\end{table}

\section{Construction of the Physical Data Precedence Table}
\label{sec:PDPT}

Priority values are compatible with all logical dependencies between quantum operations, meaning that no gate can depend, either explicitly or implicitly, on a gate with lower priority. One can therefore use the priority value to schedule the quantum gates and construct the Physical Data Precedence Table. This table has as many rows as physical qubits in the hardware and at least as many column as the maximum priority value. It indicates what physical qubits are involved in what gate at any given clock-cycle.

However, the mere knowledge of the priority values is not enough to construct the PDPT. Two non-trivial actions have to be taken in order to satisfy the connectivity constraint and resolve ambiguities for gates with equal priority: Routing operations have to be added and a tie breaking strategy needs to be introduced.

We denote with Greek letters the index of the physical qubits (corresponding to the row index for the PDPT table) and with $\tau_i$ the experimental time interval at which the gates scheduled in the $i$-th column of the PDPT are performed.

Entry $(\alpha,i)$ of the PDPT must provide two pieces of information: which logical qubit, if any, is associated with the $\alpha$-th physical qubit at clock-cycle time $\tau_i$, and what gate is currently performed on that physical qubit, if it is not idle.

\subsection{Routing operations}
\label{sec:PDPT_routing}

First of all, notice that two gates that act on (at least) one common qubit have the same priority if and only if they are both consecutive and commuting. We address this situation in the next subsection and neglect such possibility for the moment. A greedy strategy is applied when filling the PDPT starting from the highest priority gates. Let us assume that we are now scheduling gates with priority $p$. Routing operations are added at this stage through SWAP gates, the effect of a SWAP gate being to exchange the logical qubits associated with the two physical qubits involved in the SWAP.

To determine which qubits need to be exchanged, one needs to consider the ``connectivity graph'' $\mathcal{C}$ representing a sort of blueprint of the actual hardware. In fact, each node of $\mathcal{C}$ represents a physical qubit, self-loops mark the qubits where single-qubit operations are available and (undirected) edges indicate the pairs of qubits between which it is possible to implement a two-qubit gate.

When all gates with priority $(p+1)$ have been scheduled, we are left with a specific map between logical and physical qubits. We now look at the gates with priority $p$ and color the nodes of the connectivity graph $\mathcal{C}$ such that two physical qubits have the same color if and only if they are associated with logical qubits involved in the same priority-$p$ gate. No color is given to physical qubits that are idle. From the node coloring perspective, applying a SWAP gate corresponds to exchange the colors of two connected nodes. The goal is to obtain a node-coloring pattern in which the (at most two) nodes with the same color are all connected: when this is the case, all connectivity constraints for the execution of logical gates with priority $p$ are satisfied.

The development of an optimal strategy to exchange colors for the nodes in $\mathcal{C}$ is an interesting problem in itself and we indicate it as a subtask that could be optimized separately from the rest of the scheduler.  We name this sub-problem as the ``color pairing'' problem, requesting to minimize the number of color exchanges. In the following we provide a few considerations on the general case and describe an optimal (despite not unique) strategy for hardware with linear topology.

It is possible to compute a sort of distance between the current coloring pattern and an acceptable one (for which same-color nodes are connected). Given a pair of same-color nodes in $\mathcal{C}$, compute the length of the shortest path between them. Sum up all such lengths to obtain the color-pair-distance $D$. Every SWAP changes the color of at most two nodes and the color-distance can change by at most 2 (all cases may be realized, with $D$ changing by $\delta\in\{-2,-1,0,1,2\}$). We propose a heuristic that starts performing all the SWAPs leading to $\delta=-2$ and subsequently proceeds with those leading to smaller (or no) reduction of $D$. Acceptable coloring patterns are characterized by zero distance and, therefore, at least $\lceil\frac{D}{2}\rceil$ SWAPs are required to achieve it.

When $\mathcal{C}$ is a linear graph, several optimal strategies are possible. Here is one (for open boundary conditions and node index running from $0$ for the leftmost node to $N-1$ for the rightmost node):

\begin{algorithm}[H]
\caption{Color-pairing for linear graph}\label{euclid}
\begin{algorithmic}[1]
\Procedure{Left accumulation}{}
\State $n \gets 0$
\While {$n<N$}
	\If {$n$ has unique or no color}
		\State $n \gets n+1$
    \Else
    	\State find $m>n$ such that \textit{color}($n$)=\textit{color}($m$)
        \For {$k=m-1,m-2,\cdots,n+2,n+1$}
        	\State apply \text{SWAP}($k$,$k+1$)
        \EndFor
		\State $n \gets n+2$
	\EndIf
\EndWhile
\EndProcedure
\end{algorithmic}
\end{algorithm}

In Appendix~\ref{app:optimality_color_pairing_1D} we prove that such a procedure is minimal in the number of SWAP gates. However, this is not the unique optimal strategy as can be easily see by considering the symmetric procedure starting from the rightmost node. Despite multiple strategies may require the same, minimal, number of SWAPs, the final coloring pattern is different. Looking ahead to the logical gates with priority $p-1$ (and lower) may help determining which final color-pairing scheme reduces the overall routing cost. We state the extension to effective look-ahead strategies as a fascinating open problem.

\subsection{Tie breaking strategy}
\label{sec:PDPT_tiebreaking}

When gates that are both consecutive and commuting are present in the quantum circuit, then it may happen that multiple gates with equal priority $p$ act on the same qubits. A typical case is when one needs to manipulate each computational state in a coherent way according to some classical function, usually decomposed in several one- and two-qubit gates that pairwise commute. The quantum Fourier transform is another important example. Due to the exclusive activation constraint, we have to decide the order of execution. A possibility is to consider a random order. Here, we aim to do better.

To determine the order of the remaining gates and describe a general prescription, it is convenient to introduce the ``interaction graph'' $\mathcal{I}_p$. Graph $\mathcal{I}_p$ is constructed from the set of gates with priority $p$: Each node corresponds to a logical qubit, self-loops represent single qubit gates and undirected edges correspond to two-qubit gates.%

We observe that satisfying the exclusive activation corresponds to dividing the priority-$p$ gates into subsets composed by gates that act on different qubits. We mark each subset with a different color, effectively associating a color with each edge (including self loops) of the interaction graph. The problem of choosing appropriate subsets translates to the standard edge-coloring problem (no edges with the same color can share a node) in which one minimizes the number of colors involved. The fewer the colors, the larger the parallelism exploited by the scheduler. When the edges are divided in subsets, one proceeds to schedule one subset at a time while adding the routing operations according to the procedure described for gates with different priority.

However, while sets of gates with different priority are subjected to logical dependencies that pose constraint on their scheduling order, the attribution of edge colors in $\mathcal{I}_p$ is completely arbitrary. Here we propose a simple ``look ahead'' strategy to choose colors leading to a consistent logical-to-physical map between gates of multiple color subsets. This approach, which prioritizes reducing the routing cost over minimizing the number of edge colors, is expected to be advantageous for hardware with limited connectivity.

The logical-to-physical (LTP) qubit map can be seen as the identification of the nodes of the interaction graph over a subset of the nodes of the connectivity graph%
\footnote{The connectivity graph may have more nodes than $\mathcal{I}_p$, meaning that the hardware may have more physical qubits than those required by the algorithm. These additional qubits are involved in the routing operations.}.
If $\mathcal{I}_p$ is a subgraph of $\mathcal{C}$ according to the current LTP qubit map, no routing operation is required and the gates with priority $p$ can be scheduled according to the solution of the edge-coloring problem. Otherwise, one or more different LPT maps are required.

We propose to derive the new map from the solution of the maximum subgraph isomorphism problem between $\mathcal{I}_p$ and $\mathcal{C}$. Therefore one has both the initial and desired LTP map and must solve the related routing problem. When $\mathcal{I}_p$ is not fully contained in $\mathcal{C}$, the edges belonging to the maximum subgraph must be eliminated from $\mathcal{I}_p$ and a new maximum subgraph identified. Ultimately, all gates with priority $p$ will be scheduled: each group corresponding to those of a subgraph (solving edge-coloring may be required) and between them the SWAPs required by the routing.

We observe that the number of SWAP gates may be reduced by selecting the next subgraph isomorphism in ways that consider (and try to minimize) the exchange cost between the current and next LTP qubit maps.
Notice that, since finding the maximum subgraph isomorphism is a hard problem, approximate solutions are acceptable at every iteration \cite{Bahiense2012}.

\section{Scheduling QAOA for a 1D array of qubits}
\label{sec:QAOA}

We illustrate the two phases presented in the previous sections by scheduling the Quantum Approximate Optimization Algorithm (QAOA) \cite{Farhi2014,Wecker2016,Guerreschi2017a} on hardware with linear connectivity. We consider this example significant for two reasons: first, QAOA gives rise to situations where lots of commuting and consecutive gates have the same priority and the tie-breaking strategy plays a relevant role. Second, the open-boundary 1D topology reflects actual short-term devices \cite{Blatt2008,Barends2016,Harris2016} and corresponds to the most connectivity-constrained architecture that is still scalable.

QAOA is a variational algorithm to solve combinatorial problems. The quantum circuit is a sequence of only two kinds of operations, repeated for a desired number of times:
\begin{align}
\label{eq:QAOA_operations}
	\hat{U}(\gamma) &= \exp{(-i \gamma \hat{C})} \nonumber \\
	\hat{V}(\beta)  &= \exp{(-i \beta  \hat{B})} \, ,
\end{align}
with $\hat{B}=\sum_{i=0}^{N-1} \hat{X}_i$ and $\hat{C}(\hat{Z}_0,\ldots,\hat{Z}_{N-1})$, expressed in terms of the Pauli matrices $\hat{X}_i, \hat{Z}_i$ acting on the $i$-th logical qubit. $\hat{V}(\beta)$ corresponds to single-qubit rotations by the same angle on each logical qubit. $\hat{U}(\gamma)$ corresponds to a gate diagonal in the computational basis and decomposable in gates involving only $\hat{Z}$ matrices. The specific form of $\hat{C}$ depends on the problem at hand and, for the well-studied case of the MaxCut problem%
\footnote{The MaxCut problem is defined as follows: Given an undirected graph, color each node in black or white. An edge can be ``cut'' if it connects nodes with different color. Find the maximum number of edges that can be cut (providing a suitable color assignment). This is an NP-hard problem.}%
, it is the sum of parity gates like $\hat{Z}_i \hat{Z}_j$ \cite{Farhi2014,Wecker2016,Guerreschi2017a,Venturelli2018,Otterbach2017a}. One has:
\begin{align}
\label{eq:QAOA_U}
	\hat{U}(\gamma) &= \exp{\Big(-i \gamma \sum_{(i,j)\in\mathcal{I}} \hat{Z}_i \hat{Z}_j\Big)} \nonumber \\
					&= \prod_{(i,j)\in\mathcal{I}} \exp{\left(-i \gamma \hat{Z}_i \hat{Z}_j\right)} \, ,
\end{align}
where the notation $(i,j)\in\mathcal{I}$ indicates that the $(i,j)$ corresponds to an edge of the graph that defines the MaxCut instance and that also defines the interaction graph (this is the reason of the notation).

The complete quantum circuit, for a certain depth $d$, corresponds to the sequence:
\begin{align}
\label{eq:QAOA_q_circ}
	\hat{V}(\beta_{d-1}) \hat{U}(\gamma_{d-1}) \cdots \hat{V}(\beta_1) \hat{U}(\gamma_1) \, \hat{V}(\beta_0) \hat{U}(\gamma_0) \, ,
\end{align}
with $\{\gamma_k,\beta_k\}_{k=0,1,\cdots,d-1}$ being the variational parameters. The following commutation relations hold:
\begin{align}
\label{eq:QAOA_commutations}
	\left[\hat{X}_i , \hat{X}_j\right] &= 0 \nonumber \\
	\left[\hat{X}_i , \hat{Z}_j \hat{Z}_k\right] &=
    		-2i \left( \delta_{i,j} \hat{Y}_i \hat{Z}_k + \delta_{i,k} \hat{Z}_j \hat{Y}_i \right) \nonumber\\
	\left[\hat{Z}_i \hat{Z}_j , \hat{Z}_k \hat{Z}_l\right] &= 0 \, .
\end{align}
It follows that the only gates that are both commuting and contiguous belong to the same operation $\hat{U}(\gamma_k)$. The Logical Data Precedence Graph (LDPG) is straightforward to build and the priority can be assigned very easily. Noting with $t_X$ the latency of single-qubit rotations and with $t_{ZZ}$ the latency of the two-qubit parity rotations, one has that all gates $\{\exp{(-i \beta_k  \hat{X}_i)}\}_i$ have the same priority $p=(t_X + t_{ZZ})(k+1)$ and all gates $\{\exp{(-i \gamma_k  \hat{Z}_i\hat{Z}_j)}\}_{(i,j)\in\mathcal{I}}$ have priority $p=(t_X + t_{ZZ})k + t_{ZZ}$.

As anticipated, the scheduler can take advantage of the freedom to order the parity rotations composing each $\hat{U}(\gamma_k)$ by applying the strategies proposed for routing and tie breaking. We consider the case $d=1$ since a schedule for $d>1$ that requires a number of gates at most linear in $d$ is always possible. To be convinced of this fact, notice that gates $\{\exp{(-i \gamma_k  \hat{Z}_i\hat{Z}_j)}\}_{(i,j)\in\mathcal{I}}$ should be scheduled in opposite order compared of those at depth $(k-1)$ to make the logical-to-physical map of the qubits compatible between the end of $\hat{U}(\gamma_{k-1})$ and beginning of $\hat{U}(\gamma_k)$.

\begin{figure}[t!]
\centering
\includegraphics[width=1.\linewidth]{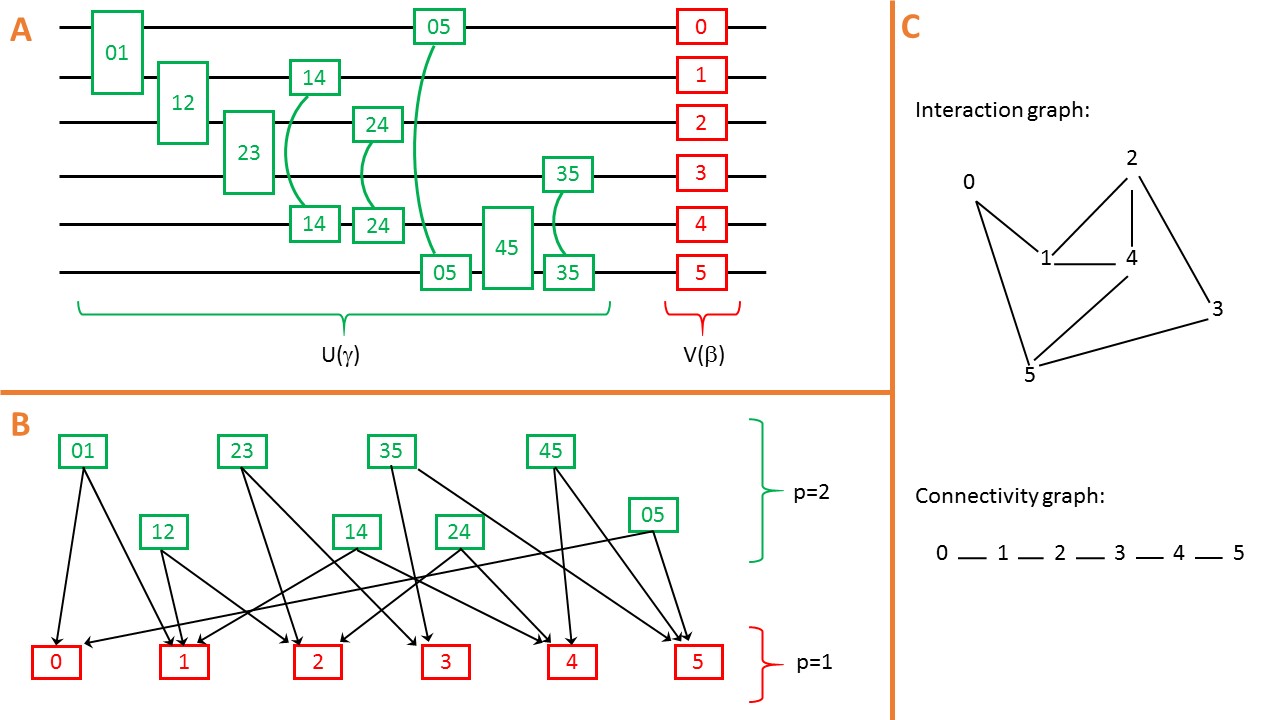}
\caption{Example of a quantum circuit for QAOA (panel A) and corresponding Logical Data Precedence Graph (panel B). All gates composing operation $\hat{U}(\gamma)$ pairwise commute, while they do not commute with the single qubit rotations composing $\hat{V}(\beta)$. The same priority is shared among the gates belonging to one of the two distinct groups. In panel C the interaction graph for priority $p=2$ is shown together with the connectivity graph.}\label{fig:QAOA}
\end{figure}

FIG.~\ref{fig:QAOA} provides a visualization of the intermediate representations discussed in the previous sections. One starts from the quantum circuit of QAOA with depth $d=1$, then constructs the LDPG and assigns a priority value to each gate. The PDPT is filled by taking into account the connectivity diagram of the hardware and, to break the ties due to $\hat{U}(\gamma)$, by exploiting approximate solutions of the maximal subgraph isomorphism with the interaction graph (here corresponding to the graph for the MaxCut instance). Due to the linear topology, the maximal subgraph isomorphism can be reformulated as looking for long paths inside $\mathcal{I}$.

In our numerical analysis, we schedule the QAOA for various problem sizes $N\in\{10, 20, 30, 40, 60, 80, 100\}$ in a way suitable for linear hardware with a number of physical qubits equal to $N$. In practice, we consider the complete utilization of the available hardware resources.
We provide the results in terms of total gate count (including the routing operations) and of the circuit depth averaged over $M=1000$ instances of MaxCut problem for random 3-regular graphs%
\footnote{A graph with undirected edges and no self-loops is $k$-regular if and only if each node has exactly $k$ edges.}%
. For simplicity, the synchronous model is considered for which $t_X=t_{ZZ}=t_{SWAP}=1$.

The schedule for the operation $\hat{V}(\beta)$ is trivial and corresponds to the same single-qubit gate $\exp{(-i \beta  \hat{X})}$ applied on each of the physical qubits, irrespective of the logical qubit associated. No routing operations are required and $\hat{V}(\beta)$ has circuit depth equal to 1. Therefore, we present our results focusing on the $\hat{U}(\gamma)$ operations where several gates are commuting and contiguous.

Three strategies are considered in increasing level of sophistication:
\begin{description}
\item[baseline] The edge-coloring problem is solved by calling the corresponding function in the Boost Graph Library \cite{2002:BGL:504206}. The initial logical-to-physical qubit map is compatible with the first edge-color. Subsequent color-pairing tasks are solved via the left accumulation strategy introduced in section~\ref{sec:PDPT_routing}.
\item[greedy] The edge-coloring problem is solved in a greedy way following a randomized sequence of the edges of the interaction graph $\mathcal{I}$. Apart from the initial logical-to-physical map, the color-pairing subtasks are solved via the left accumulation strategy. The lowest value between the baseline and $4 N$ repetitions of the greedy scheduler is used for each MaxCut instance.
\item[long-path] The edge-coloring problem is solved in two steps: First a long path (ideally a Hamiltonian path which passes through all nodes) is found in $\mathcal{I}$ and the corresponding edges colored in two, alternating, colors%
\footnote{We implemented our own code to look for long paths. In practice, it is similar to a breadth first search with the next node chosen only between those with (relative) largest connectivity.}%
. The rest of the edges are colored with a (randomized) greedy order.
It is easy to see that the initial logical-to-physical map allows the execution of all gates forming the long path without any routing, despite the fact that they have two distinct colors. The subsequent color-pairing subtasks are solved via the left accumulation strategy. The lowest value between the baseline and $4 N$ repetitions of the long-path scheduler is used for each MaxCut instance.
\end{description}

\begin{figure}[t!]
\centering
\includegraphics[width=0.82\linewidth]{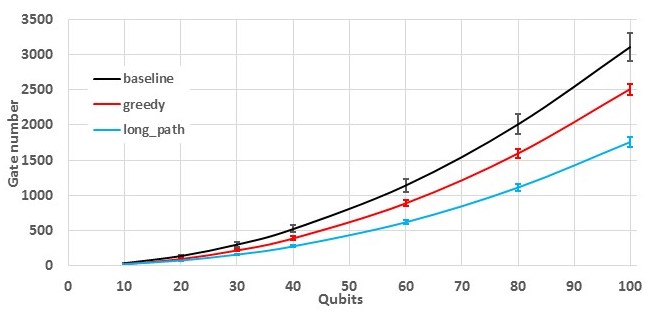}
\caption{Total number of 2-qubit gates for QAOA on linear topology as a function of the problem size $N$. Details of the three scheduling strategies are provided in the main text. Each point is obtained as the average over 1000 instances of the MaxCut problem on random 3-regular graphs and the error bar represents the standard deviation. We have considered the synchronous model having unit latency for every gate (including the SWAP operation). Including the single-qubit gates for the operation $\hat{V}(\beta)$ would add a number of gates equal to the number of qubits $N$.}\label{fig:QAOA_gates}
\end{figure}

\begin{figure}[h!]
\centering
\includegraphics[width=0.82\linewidth]{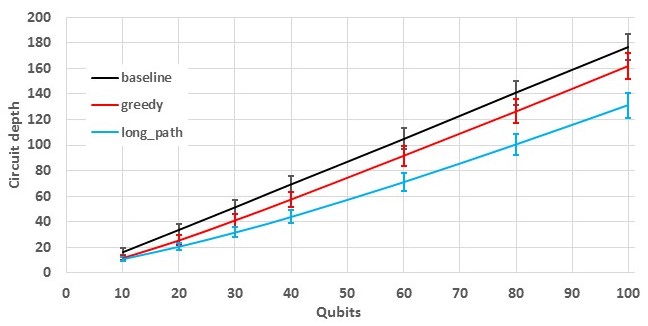}
\caption{Circuit depth for the $\hat{U}(\gamma)$ operation for QAOA on linear topology as a function of the problem size $N$. Details of the three scheduling strategies are provided in the main text. Each point is obtained as the average over 1000 instances of the MaxCut problem on random 3-regular graphs and the error bar represents the standard deviation. We have considered the synchronous model having unit latency for every gate (including the SWAP operation). Considering also the $\hat{V}(\beta)$ operation would simply increase the depth by 1, irrespective of $N$.
}\label{fig:QAOA_depth}
\end{figure}

In FIG.~\ref{fig:QAOA_gates} we report the total number of gates, including the SWAP operations, to implement $\hat{U}(\gamma)$ from the three scheduling strategies above. FIG.~\ref{fig:QAOA_depth} shows similar results for the depth of the quantum circuit. The numerical results refer to the synchronous model with unit latency per gate.


We observe that the routing overhead can be described by a quadratic function of the number of qubits $N$, as can be expected in the worst-case scenario. The circuit depth is, instead, linear in $N$ due to the possibility of performing SWAP gates in parallel. Our results suggest that, by adopting increasingly sophisticated strategies for the edge-coloring task, one obtains a consistent reduction in the number of necessary SWAP gates and in circuit depth. In Appendix~\ref{app:lower_bound_routing}, we provide additional considerations to relate our results to lower bound estimates.

\section{Conclusions}

We have presented a two-step approach to schedule quantum circuits. Three constraints have been taken into account: the logical dependency of the gates, the exclusive activation of the qubits, and their hardware-dependent connectivity. Our proposal initially captures the logical dependencies by constructing the ``logical'' data precedence graph (LDPG) and includes the routing operations in a second phase. We phrase the exclusive activation in terms of the edge-coloring problem for the interaction graph, and the connectivity constraint in terms of the color-pairing problem for the nodes of the connectivity graph.

It is important to notice that the scheduling task addressed in this study does not include all the optimizations available when compiling quantum algorithms. In fact, we have supposed that the quantum algorithm is initially provided as a sequence of one- and two-qubit gates that are readily available in the hardware of interest. In general, this sequence is the output of another compilation task that may perform all or a subset of the following optimizations:
decompose an arbitrary single-qubit rotation as a sequence of fixed-angle rotations, possibly introducing a controlled approximation on the actual rotation angle; decompose multi-qubit gates into a sequence of one- and two-qubit gates; combine sequences of operations into a smaller number of gates (for example an arbitrary long sequence of single-qubit rotation can be compressed to only three rotations using the Euler decomposition of tridimensional rotations); attempt to exchange the order of logical operations (even when the corresponding quantum gates do not commute, this might still be possible by properly modifying one or both of the gates) to trigger further simplifications.

Within the scope of our work, the main limitation of the two-step approach is the lack of a ``look-ahead'' strategy that may cause the routing at priority $p$ to be undone (paying the overhead cost) at priority $p-1$ or lower. We suggest a way to mitigate such effect through the approximated solution of the maximal subgraph isomorphism problem between the connectivity and interaction graphs. The expectations are confirmed by the numerics related to the QAOA. In fact, the long-path strategy is the specialization of the maximal subgraph isomorphism for linear connectivity graphs.

Quantum computing has only recently reached technological relevance, but the need and demand of effective solutions for all the ancillary tasks required by the practical realization of quantum algorithms is growing and already strong. We believe that gate scheduling represents one of the most prominent tasks to solve to fully take advantage of quantum speedups, and have proposed an effective approach to address it.

\section*{Acknowledgments}
The authors would like to thank Mikhail Smelyanskiy for posing the question at the origin of this study and Nicolas Sawaya for helpful comments on the manuscript.




\appendix
\renewcommand\thefigure{\thesection.\arabic{figure}}
\renewcommand\thetable{\thesection.\arabic{table}}
\setcounter{figure}{0}
\setcounter{table}{0}

\bigskip\noindent\makebox[\linewidth]{\resizebox{0.3333\linewidth}{1pt}{$\bullet$}}\bigskip

\section{Proof of optimality of left accumulation for color-pairing in 1D}
\label{app:optimality_color_pairing_1D}

This appendix demonstrates that the ``left accumulation'' strategy presented in section~\ref{sec:PDPT} is optimal for the color-pairing problem in one-dimensional topology with open boundary conditions, meaning that the physical qubits are disposed on a line with the end qubits not connected to form a ring.
Notice that this strategy is not unique (consider for example the symmetric strategy of ``right accumulation'') and that the overall task of minimizing the routing cost of scheduling a quantum algorithm depends on how the color-pairing strategy of gates with priority $p$ influences the routing cost of those with priority $p-1$. It may be possible that solving a subtask according to a locally suboptimal strategy leads to globally more efficient schedule for certain problem instances.

The color-pairing problem can be stated as follows:
Given a connectivity graph $\mathcal{C}$, consider a (node) coloring of its nodes such that at most two nodes share the same color. It is possible to exchange the color of two connected nodes (this operation corresponds to a SWAP gate). The goal is to reach a compatible color pattern, \emph{i.e.} one in which all nodes sharing the same color are connected, with the minimum number of exchange operations.

Recall the definition of the color-pair distance when $\mathcal{C}$ is a line: The color-pair distance $D$ is the sum of the number of nodes separating any pair of same-color nodes.
Since every SWAP exchanges the color of at most two nodes and the color-distance can vary by at most 2 (all cases may be realized, with $D$ changing by $\delta\in\{-2,-1,0,1,2\}$). Acceptable coloring patterns are characterized by a zero distance and, therefore, at least $\lceil\frac{D}{2}\rceil$ SWAPs are required to achieve it.

For the purpose of color pairing, nodes with unique colors can be safely assumed to be colorless. In the optimality proof, we consider such situation separately. Observe that a single exchange operation has only six different outcomes. When only one node is colored (either
\tikz\draw[black,fill=blue] (0,0) circle (4pt);
\tikz\draw[black,fill=white] (0,0) circle (4pt);
or
\tikz\draw[black,fill=white] (0,0) circle (4pt);
\tikz\draw[black,fill=blue] (0,0) circle (4pt);
):
\setlist{nolistsep}
\begin{enumerate}[noitemsep]
\item{} Moving the colored node closer to its companion reduces $D\rightarrow D-1$.
\item{} Moving the colored node farther away from its companion increases $D\rightarrow D+1$.
\end{enumerate}
When both nodes have the same color (as for
\tikz\draw[black,fill=blue] (0,0) circle (4pt);
\tikz\draw[black,fill=blue] (0,0) circle (4pt);
):
\setlist{nolistsep}
\begin{enumerate}[noitemsep]
\setcounter{enumi}{2}
  \item{} No changes concerning the color pairing or $D$.
\end{enumerate}
When the two nodes have different color (either
\tikz\draw[black,fill=blue] (0,0) circle (4pt);
\tikz\draw[black,fill=red] (0,0) circle (4pt);
or
\tikz\draw[black,fill=red] (0,0) circle (4pt);
\tikz\draw[black,fill=blue] (0,0) circle (4pt);
):
\setlist{nolistsep}
\begin{enumerate}[noitemsep]
\setcounter{enumi}{3}
  \item{} The ``locally best'' move brings both same-color node pairs closer to each other. The distance diminishes by one for each color and then $D\rightarrow D-2$.
  \item{} The ``locally balanced'' move brings one same-color pair closer while separate the other pair farther apart. As a consequence, $D$ is unchanged.
  \item{} The ``locally worst'' move brings both same-color node pairs farther away from each other. The distance increases by one for each color and then $D\rightarrow D+2$.
\end{enumerate}

It is clear that any optimal strategy must not include any move of type-2, 3, or 6.
When possible, moves of type-4 are preferable over those of type-1 and 5. The best scenario is when moves of type-4 suffice to obtain a compatible color pattern. However, moves of type-1 are unavoidable if there is a colorless node between two same-color ones. In particular, if a pair is separated by $k$ colorless nodes, the number of required moves of type-1 is also $k$ (an example with $k=2$ is
\tikz\draw[black,fill=blue] (0,0) circle (4pt);
\tikz\draw[black,fill=white] (0,0) circle (4pt);
\tikz\draw[black,fill=white] (0,0) circle (4pt);
\tikz\draw[black,fill=blue] (0,0) circle (4pt);
). Of course, the same colorless node may separate multiple same-color nodes and therefore contribute a type-1 move for each such pair.

Are there situations when moves of type-5 are unavoidable? Yes, it happens when two same-color nodes are separating another pair (think of the sequence
\tikz\draw[black,fill=blue] (0,0) circle (4pt);
\tikz\draw[black,fill=red] (0,0) circle (4pt);
\tikz\draw[black,fill=red] (0,0) circle (4pt);
\tikz\draw[black,fill=blue] (0,0) circle (4pt);
). To achieve one of the two compatible color patterns (either
\tikz\draw[black,fill=blue] (0,0) circle (4pt);
\tikz\draw[black,fill=blue] (0,0) circle (4pt);
\tikz\draw[black,fill=red] (0,0) circle (4pt);
\tikz\draw[black,fill=red] (0,0) circle (4pt);
or
\tikz\draw[black,fill=red] (0,0) circle (4pt);
\tikz\draw[black,fill=red] (0,0) circle (4pt);
\tikz\draw[black,fill=blue] (0,0) circle (4pt);
\tikz\draw[black,fill=blue] (0,0) circle (4pt);
) one needs a type-5 move followed by a type-4 move. Any optimal strategy thus require one type-5 move for every same-color pair contained between, and separating, another same-color pair.

It is straightforward to verify that the left accumulation strategy presented in section~\ref{sec:PDPT} avoids any move of type-2, 3, and 6. In addition, it requires the minimum number of type-1 and 5 moves. The optimality follows logically.

\section{Lower bound for the number of routing operations}
\label{app:lower_bound_routing}

The numerical study in section~\ref{sec:QAOA} shows that adopting our two step approach provides encouraging results for the problem of scheduling quantum algorithms on hardware with linear connectivity. The efficacy of the scheduler is improved by adding stochasticity and by initializing the logical-to-physical qubit map according to long paths in the interaction graph.
An important question remains unanswered: How close are the schedules we found compared to the globally-best schedule?

In this section, with globally-best schedule we refer to the schedule involving the least number of gates, without considering the circuit depth. The two quantities are clearly related, but not always the circuit with fewer gates results in the shallowest depth. We consider two approaches: The first is the exhaustive enumeration of all the possible circuits, while the second is a heuristic bound.

\subsection{Exhaustive search}
\label{app:exhaustice_search}

This approach is only feasible for extremely small problem sizes. The number of quantum circuits involving $N$ qubits and including at most $S$ exchange operations grows according to:
\begin{equation}
	\left| \{ \text{initial LTP maps} \} \right| \times
	\left| \{ \text{SWAP sequences of length }k \} \right|
    \leq N! \times (N-1)^S
\end{equation}
There are obvious situations that can be excluded from the search: For example, starting from an initial LTP map or its reverse order are equivalent situations (reducing the first term in the expression above by a factor 2) and eliminating situations in which the same SWAP gate is applied twice consecutively can be proved not to eliminate uniquely optimal strategies%
\footnote{This claim is more complex than just stating that two identical consecutive SWAPs cancel each other. In fact, after the first SWAP, up to two gates $\exp{(-i \gamma  \hat{Z}_i\hat{Z}_j)}$ might become possible that were between previously unconnected qubits. However one can find modified schedules with the same number of SWAPs that satisfy the constraint.}.

The number of possible quantum circuits is reduced to $\frac{N!}{2} (N-1)(N-2)^{S-1}$. Unfortunately, its scaling is still very unfavorable and, in practice, this limits our current numerical results to $N\leq 8$. For $N=8$, we solved all 150 instances considered by exploring $S\leq 9$. To allow such broad exhaustive search, we further reduce the number of the SWAP sequences explored by following the method below.
We observe that each exchange sequence can be thought as a number with $S$ digits in base $(N-1)$, where $(N-1)$ are the physically distinct SWAP gates available for a line of length $N$. We order the SWAP sequences in increasing order, with the most significant digit identifying the first SWAP and the least significant digit identifying the last SWAP, and start evaluating them once at a time.
For each specific sequence, we compute how many logical gates are left unscheduled since they involve logical qubits that never became adjacent. Since any exchange operation can modify the connectivity between logical qubits in a way that at most two additional pairs of qubits become connected due to that particular SWAP gate, we can deduce the minimum number of changes in the SWAP sequence that may allow a solution. For example, if 3 logical gates are not possible with a specific SWAP sequence, at least two final SWAPs have to change. In general, if $r$ gates are not possible, then at least $\lceil \frac{r}{2} \rceil$ SWAPs at the end of the sequence have to change.
We report the values found for the total number of gates for a single $\hat{U}(\gamma)$ operation in Table~\ref{table:exhaustive_search}.

\begin{table}[ht]
\centering
\begin{tabular}{ | c || c | c | c |}
\hline
    \hspace{4mm}$N$\hspace{4mm} & \hspace{3mm}exhaustive\hspace{3mm} &      greedy       &     long path \\
                                &   search   & (500 repetitions) & (500 repetitions) \\
\hline \hline
     4 & 3    $\pm$ 0    &  3    $\pm$ 0    &  3    $\pm$ 0    \\ \hline
     6 & 5.11 $\pm$ 0.32 &  5.96 $\pm$ 0.58 &  6.11 $\pm$ 0.33 \\ \hline
     8 & 7.5  $\pm$ 1.09 &  9.53 $\pm$ 1.29 &  9.19 $\pm$ 1.17 \\ \hline
    10 &                 & 14.51 $\pm$ 1.80 & 12.44 $\pm$ 1.94 \\ \hline
    12 &                 & 21.51 $\pm$ 3.04 & 17.45 $\pm$ 2.79 \\ \hline
\end{tabular}
\vspace{0.3cm}
\caption{Number of SWAP gates to schedule QAOA on a line of $N$ qubits; results averaged over 150 instances. In the exhaustive search with $N=8$, all 150 instances were solved with sequences of at most $S=9$ exchange operations.}
\label{table:exhaustive_search}
\end{table}

\subsection{Heuristic lower bound}
\label{app:heuristic_lower_bound}

It is possible to provide a lower bound on the number of SWAP operations required to schedule all terms of $\hat{U}(\gamma)$ that is based on the adjacency matrix of the interaction graph $\mathcal{I}$.

Let us fix the initial LTP qubit map, so that the adjacency matrix is uniquely defined.
Due to the particular properties of linear connectivity, the distance from the diagonal (row-wise or column-wise does not matter because the interaction graph is undirected thus its adjacency matrix is symmetric) of every non-zero entry corresponds to the number of SWAP operations minus 1 required to move the two qubits involved in the corresponding gate in contact. For example, consider a specific non-zero entry in position $(i,j)$. For the adjacency matrix it means that the logical qubit mapped to physical qubit $i$ needs to interact with the logical qubit mapped to physical qubit $j$. One needs $|i-j|-1$ SWAPs to move the logical qubits along the line until they are adjacent.
If for each row we consider only the non-zero entry that is the most distant from the diagonal, we are effectively relaxing the problem. Let us denote with $W$ the quantity obtained by summing up the distances of all these entries.
It is tempting to consider this a lower bound for the number of SWAP gates required to implement $\hat{U}(\gamma)$ given the initial LTP map. This is not strictly correct since $W$ must be:
\begin{itemize}
\item  divided by 2 since every gate corresponds to two non-zero entries and they may contribute to two different rows. Specifically, non-zero entry $(i,j)$ has its symmetric non-zero entry in $(j,i)$. Both entries represent the same logical gate, \emph{i.e.} edge of the interaction graph, but contribute to both row $i$ and $j$.
\item  divided by $2\,k$, with $k$ being the maximum degree of the interaction graph, since a single SWAP exchange two columns (and rows) affecting at most $2\,k$ non-zero entries and possibly brings each of them closer to the diagonal by one position.
For example, consider two non-zero entries at position $(i,j)$ and $(h,j)$ such that $i,h>j+1$, \emph{i.e.} the entries are in the same column and below the diagonal. Swapping two columns $i$ and $i+1$ moves both entries closer to the diagonal to position $(i,j+1)$ and $(h,j+1)$.
A similar effect may involves also entries originally in column $(j+1)$. Due to the degree of connectivity, each column has at most $k$ non-zero entries.
\end{itemize}
For the class of instances considered, 3-regular random graphs have $k=3$.

Finally, we have to relax the constraint of having a fixed LTP map. This can be done by considering each permutation of the rows and columns of the adjacency matrix. The scope is searching for the LTP map that reduces the profile of the adjacency matrix \cite{Norman1976}, effectively providing the minimum value of the quantity $W$.

So far our considerations are exact, but minimizing the profile is a NP-hard problem \cite{Garey1990}. We estimate the profile by using a heuristic method based on the reverse Cuthill-McKee (RCM) algorithm \cite{Cuthill1969}.

\begin{figure}[bht]
\centering
\includegraphics[width=0.8\linewidth]{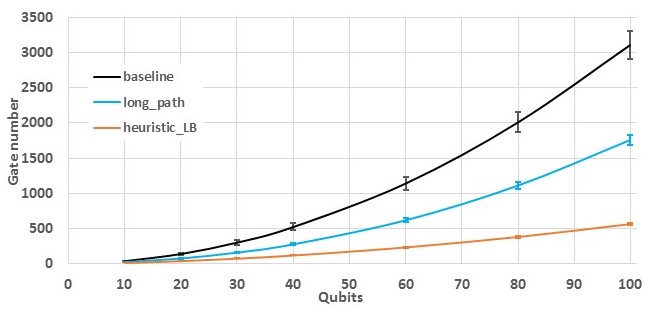}
\caption{
Total number of 2-qubit gates for QAOA on linear topology as a function of the problem size $N$. Each point is obtain as the average over 1000 instances of the MaxCut problem on random 3-regular graphs and the error bar represents the standard deviation. We have considered the synchronous model having unit latency for every gate (including the SWAP operation). The baseline and ``long path'' approaches are described in section~\ref{sec:QAOA}. The heuristic lower bound is obtained as described in Appendix~\ref{app:heuristic_lower_bound}: Most probably it undercounts the number of necessary gates, but it is not mathematically guaranteed to be lower than the actual minimum.
}
\label{fig:heuristic_lower_bound}
\end{figure}

In Figure~\ref{fig:heuristic_lower_bound}, we report the heuristic lower bound of the number of gates to implement a single operation $\hat{U}(\gamma)$ computed as the minimum profile (minus $N$) of the adjacency matrix divided by 12, plus the number of two-qubit parity rotations (those are in number $3N/2$). Observe that, while the heuristic estimate probably underestimates the number of SWAP operations, the fact that the solution of the minimum average bandwidth is approximate does not allow us to claim a rigorous lower bound.

To understand why we expect the heuristic lower bound to undercount the number of SWAP gates, consider the fact that we divide the average bandwidth by $2k$. In the language of Appendix~\ref{app:optimality_color_pairing_1D}, this means that only type-4 moves are considered, \emph{i.e.} those SWAPs that reduce the color-pair distance $D$ by 2. In addition, we also consider that each exchange operation counts as a type-4 move with respect to color-pairing distance for all future logical gates involving one of the exchanged qubits. It would not be surprising for the heuristic lower bound to be, for example, a factor 2 smaller than the actual minimum, at least for large enough systems. To address the performance of our scheduling methods, it would be interesting to have access to a tighter estimate of the lower bound.

\bibliographystyle{apsrev4-1}
\bibliographystyle{unsrt}
\bibliography{q_compiler_1D}


\end{document}